\documentclass[11pt,a4paper]{article}

\usepackage{graphicx}

\newcommand{\UU}{{\bf U }}
\newcommand{\uu}{{\bf u }}
\newcommand{\lb}{{\bf l }}

\newcommand{\ff}{{\bf f }}
\newcommand{\xx}{{\bf x }}
\newcommand{\kk}{{\bf k }}
\newcommand{\pp}{{\bf p }}
\newcommand{\rr}{{\bf r }}
\newcommand{\vv}{{\bf v }}
\newcommand{\qq}{{\bf q }}

\begin{document}

\title{A {LES}-Langevin model for turbulence}

\author{J.-P. Laval$^{1}$, B. Dubrulle$^{2}$\\
\\
$^{1}$ Laboratoire de M\'ecanique de Lille, CNRS, UMR 8107,Blv Paul Langevin,\\
 F-59655 Villeneuve d'Ascq Cedex, France\\
\\
$^{2}$ Groupe Instabilit\'es et Turbulence, \\
CNRS, URA 2464, SPEC/DRECAM/DSM/CEA Saclay,\\
F-91191 Gif sur Yvette, Cedex}

\date{\today}
\maketitle

\abstract{
We propose a new model of turbulence for use in large-eddy
simulations (LES). The turbulent force, represented here by the
turbulent Lamb vector, is divided in two contributions. The
contribution including only subfilter fields is deterministically
modeled through a classical eddy-viscosity. The other contribution
including both filtered and subfilter scales is dynamically computed 
as solution of a generalized (stochastic)  Langevin equation. This 
equation is derived using Rapid Distortion Theory (RDT) applied to 
the subfilter scales. The general friction operator therefore 
includes both advection and stretching by the resolved scale.
 The stochastic noise is derived as the sum of a contribution
from the energy cascade and a contribution from the pressure. The LES model is
thus made of an equation for the resolved scale, including the
turbulent force, and a generalized Langevin equation
integrated on a twice-finer grid. We compare the full model with
 several approximations. In the first one, the friction operator 
of the Langevin equation is simply replaced by an empirical
constant, of the order of the resolved scale correlation time. In the
second approximation, the integration is replaced by a condition of
instantaneous adjustment to the stochastic force. In this
approximation, our model becomes equivalent to the
velocity-estimation model of Domaradzki et al.
\cite{domaradzki97,domaradzki99,domaradzki00}. In the isotropic,
homogeneous situations we study, both approximations provide
satisfactory results, at a reduced computational cost.
The model is finally validated by comparison to DNS and is tested
against classical LES models for isotropic homogeneous turbulence,
based on eddy viscosity. We show that even in this situation, where
no walls are present, our inclusion of backscatter through the
Langevin equation results in a better description of the flow.


\section{Introduction}
The rationale for Large Eddy Simulation is often rooted in our
inability to handle all the degrees of freedom of a large Reynolds
number turbulent flow. Given that the smaller scales monopolize most of
the computing resources, it is tempting to cut through the
number of degrees of freedom via an ad hoc small-scale decimation. The
price to pay is of course a need for parameterization (the so-called
SubGrid Scale models, or SGS), to make up for
the energy transfer of the ghost scales. We refer the reader to
\cite{domaradzki02,pope00,meneveau00,piomelli99} for
recent reviews about modern parameterization strategies
in LES. We may loosely divide the SGS models in two classes,
according to their philosophy: the "deterministic" models based on
eddy-viscosities, and the "stochastic" models, based on synthetic
fields. In eddy-viscosity methods, the action of small scales is parameterized via a few
deterministic numbers, linked with the various components of the
subgrid-scale stress tensor. These models seek to reproduce the
intensification of energy transport due to the action of scales
widely separated from the considered one. However, they fail to
reproduce backward energy transfer (backscatter) from small to large
scale, created by elongated triads in the spectral
space\cite{ossia00}. This effect has been shown to induce a
stochastic character in the LES \cite{leith90}. In ideal situations, where
the turbulence is isotropic, homogeneous and far from wall, this
backscatter is usually viewed as secondary, and eddy-viscosity based
models are generally satisfactory. However, in more realistic
situations, including turbulence near walls or in boundary layers, the
energy backscatter has proven to be an essential
feature\cite{clark79,meneveau94}. Some models like mixed models
based on similarity hypothesis are able to reproduce a realistic 
backscatter in some situations (\cite{horiuti97,liu94}).
The need for backscatter modeling also leads to the development 
of "stochastic" strategies, where the discarded small-scale motions 
are replaced by a set of random numbers, mimicking either a random force
or synthetic velocity fields \cite{domaradzki97,scotti99,laval03a}.
In many ways, these strategies resemble the strategies used to
describe the dynamics of a heavy particle coupled to a thermal bath
involving many degrees of freedom (the so-called Brownian motion).
The decimation is here performed by substituting in place of the bath
a deterministic friction and a stochastic force, the two terms being
linked through the dissipation theorem. The initial problem is then
completely described through the so-called Langevin equation.\

Turbulence is typically an out-of-equilibrium system, and there is
probably no hope that such a simple description will ever be possible
(would it be only because no fluctuation-dissipation theorem prevails
in turbulence!). However, we would like to use this analogy to
motivate a new strategy for LES modeling: replace the actual dynamics
of the decimated degrees of freedom by a suitable noise, via a
Langevin equation. Although  this strategy may seem close
to recent models based on synthetic fields, we would like to point
out an important philosophical difference: rather than trying to
estimate {\sl the actual} small-scale dynamics, we aim at trying to
estimate {\sl a plausible} small-scale dynamics. We believe there is no 
unique solution for this last option. In the sequel, we present one 
solution based upon Rapid Distortion Theory \cite{laval03a}. There 
may exist actually more efficient models, based e.g. upon information 
theory \cite{jaynes57}.\

To be more specific, consider a turbulent flow, with velocity field
$u_i(x,t)$ and introduce a filtering procedure so as to separate it into a
resolved field $U_i=\overline{u_i}$ and a subfilter field
$u_i^\prime=u_i-U_i$. The resolved field obeys a dynamical equation
obtained by filtering of the Navier-Stokes equation, which may
conveniently be written as \cite{wu99}:
\begin{eqnarray}
&&\partial_t \UU + \overline{( \UU\cdot \nabla) \UU}  +
\overline{\lb}  + \overline{(\uu^{\prime} \cdot\nabla)
\uu^{\prime}} = -\nabla P+\nu \Delta \UU,\nonumber\\
&&\lb=(\UU\cdot\nabla) \uu^{\prime}+(\uu^{\prime} \cdot\nabla)\UU.
\label{grande}
\end{eqnarray}
Here, $P$ is the resolved pressure and $\nu$ is the viscosity and $\lb$ is
a « turbulent force ». In idealistic situations including a spectral
 gap between resolved and subfilter scale, the vector $\lb$ is zero,
 and one can rigorously show that the contribution of the 
remaining term is of "diffusive" type (providing
certain symmetries which exclude first order behavior such as the
AKA effect \cite{dubrulle90}). Even as one departs from this
idealistic situation, experimental \cite{tong98} and numerical study
\cite{laval01a} show that this term correlates strongly to the resolved
velocity gradient, thereby allowing a deterministic treatment through
an eddy-viscosity of appropriate shape. In the same time, the force
vector $\lb$ becomes increasingly non-negligible (it can even become
dominant in 2D situations see \cite{laval03a}). It is responsible for
backscatter type of behaviour and needs to be modeled through novel
"non-diffusive" and "non-deterministic" strategies.
In the sequel, we will therefore focus onto the modeling of the
$\lb$ term, via a generalized Langevin equation
${\partial}_t\lb=A\lb+{\bf \xi}$ where $A$ is a generalized evolution
operator, and $\xi$ is a noise.\

However, a clear difficulty associated with this strategy is the lack
of theoretical guide (equivalent of the statistical mechanics in
Brownian motion) to help us devising the "best" friction, and the
"best" stochastic force. For the time being, we then choose to pin
down our model as best as possible to the real dynamics of
Navier-Stokes by trying to derive it from the
original dynamical equations, rather than from pure
empirical or dimensional considerations. For this, we reformulate
the RDT-Langevin model of Laval et al. \cite{laval03a} in a way
suitable for LES. There are of course limitations to this approach,
pertaining the need for both a simple enough model, and  for
tractable analytical computations. We try to formulate them as
honestly as possible by pointing out the approximation we make at the
various stage of the derivation of the model. This is done in Section
~\ref{sec:RDT}. The
LES-Langevin model is then implemented in the case of
periodic three-dimensional turbulence and comparison
with other existing LES models is provided in Section~\ref{sec:VN}. Our
conclusions follow in Section~\ref{sec:CONCL}.

\section{The LES-Langevin model of turbulence}
\label{sec:RDT}

\subsection{Derivation of the Langevin equation}
Our derivation is based on the stochastic RDT model developed by
Laval {\sl et al.} \cite{laval04a,dubrulle04a,laval03a}.
This model is based on the observation that
subfilter-scales are mostly slaved to resolved scales via linear processes
akin to rapid distortion. This property is substantiated by various
numerical simulations, and is linked with the prominence of non-local
interactions at subfilter-scale \cite{laval03a}. Using incompressibility, the
small-scale dynamics in this model can be written as:
\begin{equation}
\partial_t \uu^\prime=-\lb^\prime -\nabla p^\prime+\nabla(\nu+ \nu_t) \nabla
\uu^\prime-{\bf f}.
\label{eq:SRDT}
\end{equation}
Here, $p^\prime$ is the subfilter pressure,
$\nu_t$ is a turbulent viscosity describing the
non-linear
interactions between subfilter scales and ${\bf f}$ is a forcing
stemming from the energy cascade. The latter can be shown to be
dominated by resolved scales non-linearities
through $f_i=\partial_j\left(U_i U_j-\overline{U_i U_j}\right)$.
Finally,
we may use the observation that subfilter scales vary over fast time
scale with respect to resolved scales to write:
\begin{eqnarray}
\partial_t \lb &\approx & ({\bf \UU} \cdot\nabla)\partial_t
\uu^{\prime}+(\partial_t {\bf \uu}^{\prime} \cdot\nabla) \UU,\nonumber\\
    &\approx & - \left\{ ( {\bf \UU} \cdot\nabla) (\lb^\prime +
    \ff)) + \left[(\lb^\prime +
\ff)\cdot\nabla\right]\UU \right\} \nonumber\\
&&- \left\{ ( {\bf \UU} \cdot\nabla)\nabla
p^\prime + \left[(\nabla
p^\prime)\cdot\nabla\right]\UU \right\}
\nonumber\\
&&+ visc,
\label{eq:lamb}
\end{eqnarray}
where $visc$ gathers all the term containing $\nu$ or $\nu_t$.
This equation is only an approximation, in so far as the assumption
of "rapidly varying scales" becomes less and less valid as the
resolved scales and subfilter velocities are becoming closer and closer in
scale space. However, it seems to capture the
dominant physics of the evolution of the vector
$\lb$, as will be later shown. Further, it is tempting to 
simplify the viscous terms of the second equation
of (\ref{eq:lamb}) to try and get a closed equation for $\lb$.
Indeed, these  terms involve an a priori rather arbitrary
turbulent viscosity and one could redefine it so that the viscous
terms are simply lumped into a term $\nabla
(\nu+\nu_t)\nabla \lb$. Finally, we note that the
terms involving the subfilter pressure depend on
boundary conditions and on subfilter velocities,
so that they can be thought to vary over a fast
time scale, contrarily to the $\ff$ term which
varies over a slow time scale. We therefore
choose to collect all the term involving the
pressure and $\ff$ into a « noise term », ${\bf
\xi_0+\xi}$, such that ${\bf \xi}$ is a Gaussian-centered noise and :
\begin{eqnarray}
&&\xi_0=-(\UU\cdot\nabla)
\ff-(\bf \ff\cdot\nabla)\UU,\nonumber\\
&&<{\bf \xi}>=0, \nonumber\\
&& <\xi_i(t,x)\xi_j(t',x')>= T_{ij}(t-t',x-x'),
\label{noise}
\end{eqnarray}
where $T_{ij}$ is the noise correlation function,
to be specified. Note that since $\xi$ comes from
a pressure contribution, it only affects the
non-soleinodal part of the turbulent force. In
the sequel, since we work with periodic boundary
conditions, we can apply a simple projection
procedure to work only with soleinodal fields,
thereby discarding the $\xi$ term. We therefore
leave the study of the influence of this term for
future work, involving more realistic boundary
conditions such as flow near walls.\

Collecting all the results, we therefore obtain the following RDT
based-model for the turbulent force $\lb$ as:
\begin{equation}
\partial_t \lb =-(\UU\cdot\nabla)
\lb^\prime -(\lb^\prime \cdot\nabla)\UU +\nabla(\nu + \nu_t) \nabla
\lb+{\bf \xi_0+\xi}
\label{eq:modelRDT}
\end{equation}
It takes the form of a generalized Langevin
equation, with friction made of viscosity and rapid distortion by
resolved scales, and with stochastic forcing, with mean value generated through
the energy cascade, and whose correlations are
physically imposed by boundary conditions via
pressure terms.

Before implementing this Langevin equation into a LES model, we first
validate it through dynamical a priori tests
and seek optimal performances by tuning of $\nu_t$. Our
LES-Langevin model will follow after this validation step.

\subsection{Dynamical a priori testing}

\subsubsection{The numerical procedure}

The numerical simulations described in the present paper have been
performed with a spectral code over a cubic domain. This choice is
mainly motivated by its performances in terms of accuracy and its
versatility regarding any variant of our model. The aliasing is
removed by keeping only the $2/3$ largest modes in each direction.
The time integration is performed with a second order Adams Bashforth
scheme. The simulations with both the  RDT and the LES-Langevin
models are performed with a sharp cut-off filter
at $k=k_c$. The filter may not be optimum in terms of efficiency, but it has the
interesting properties to clearly separate scales and to commute
with derivative operators. The influence of the filter would 
need to be conducted for a complete
validation. The validation tests are performed on decaying and forced
isotropic homogeneous turbulence. The simulation of decaying turbulence is
initialized with a Gaussian velocity field with a given energy spectra
$E(k,t_o) \propto  k^4 exp(-k^2/8)$. For the forced simulation, the
forcing {\bf F} is defined by ${\bf F}(\kk,t) \, dt = \gamma(k,t) \, {\bf u}(\kk,t)$,
with $\gamma$ chosen so that the difference of the energy spectra before and after
the forcing (i.e. the energy injection rate $\Im$) is constant in
time ($\Im=1.$). The forcing is concentrated at the lowest
wavenumbers $\kk$ such as $0<k<1$. The initial time of the forced simulation is
chosen so that  statistical properties are stabilized.
The simulations are integrated over approximately 6.3 turnover time in
the decaying case and 4 turnover time (after stabilization) in the forced case. A
constant molecular viscosity of $\nu=0.0014$ is used for all simulations. The resulting
Taylor Reynolds number is approximately constant and equal to
$R_\lambda=200$ for the forced DNS and varies from $R_\lambda=260$ to $R_\lambda=26$
in the decaying case. Both DNS are performed with $341^3$
effective wavenumbers after desaliasing ($512^3$ grid points). The space
resolution allows a computation of scales down to 0.5 and 0.8
Kolmogorov scale for the decaying and the forced DNS respectively.
 The simulation with LES-Langevin model are performed with $k_c=21$ 
($42^3$  effective wavenumbers are used for the resolved scales).

\subsubsection{Validation of the model}
The validation of the model is conducted by dynamical tests performed
with the DNS of decaying 3D turbulence. It is performed by comparing
a full DNS   and a simulation at the same
resolution, in which the turbulent force $\lb$ is
replaced by the solution of (\ref{eq:modelRDT}).
This way, we may explore the validity of the approximations we made in its
derivation. We focus here on the energy spectra, so as
to capture possible deficiency of our model regarding energy transfer between
scales. The test is performed for  $1<t<2$
(all the simulations are initialized with the same field at t=1).
By that time, the kinetic energy is divided by a factor of approximately 2.
The energy spectra at time $t=2$ are shown in 
fig. \ref{fig:DNS-M047-M048-M046-decaying}.

\begin{figure}[h]
\centerline{\includegraphics[width=0.75\columnwidth]{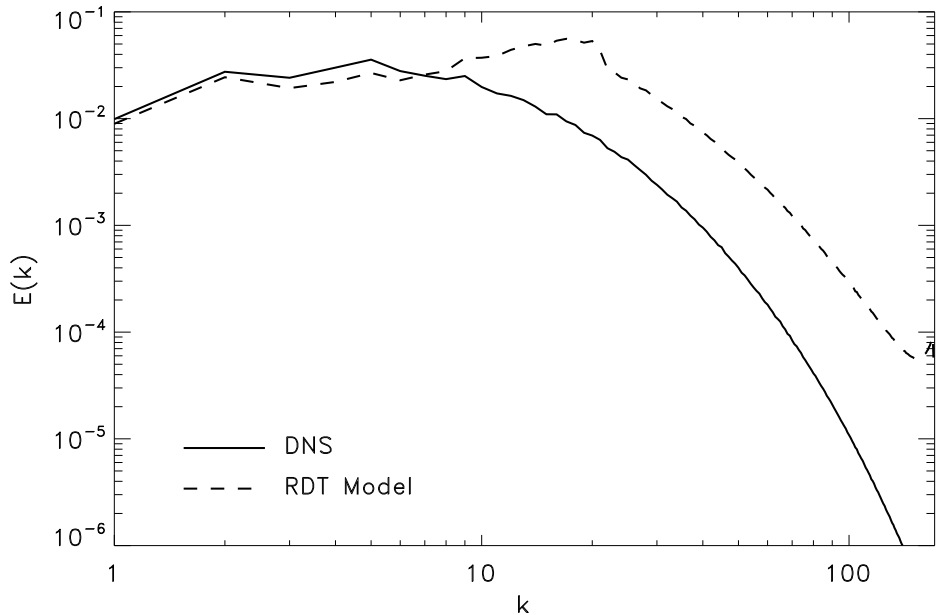}}
\caption{ Energy spectra of the
RDT Models (\ref{eq:modelRDT}) and the 
equivalent DNS at $t=2$  of simulations of
decaying turbulence (the RDT simulation is 
initialized with the velocity field of the DNS at t=1)}
\label{fig:DNS-M047-M048-M046-decaying}
\end{figure}

The model (RDT) and the DNS agree at the largest scales ($k<3$) but they
significantly differ for smaller scales (the bump at k=21
is due to the coupling between the equation for resolved scales and the
equation for subfilter scales running in parallel).
This can be explained through the analysis of the
evolution of $\lb$ (fig. \ref{fig:Lamb-M046-decaying}).
\begin{figure}
\centerline{\includegraphics[width=0.75\columnwidth]{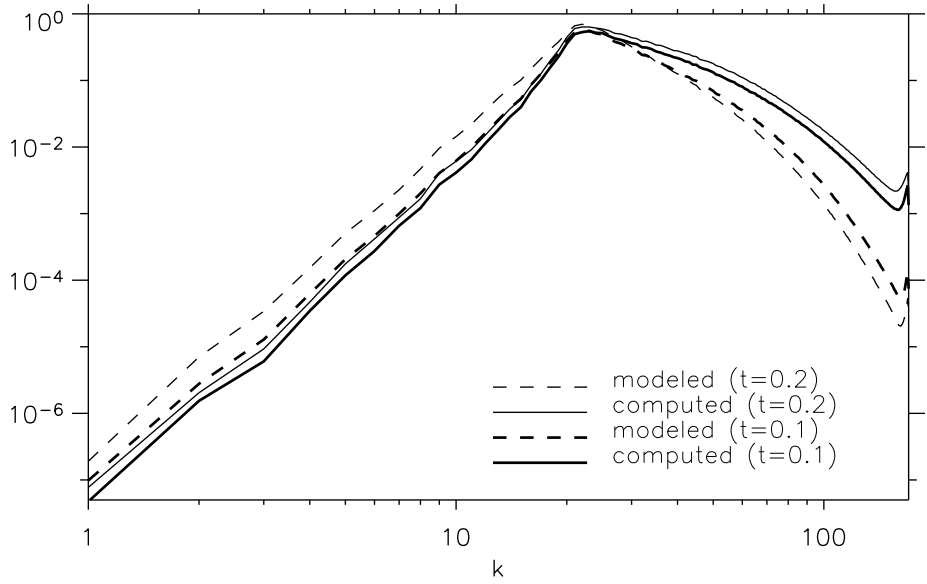}}
\caption{ Comparison at two different times
of the spectral density of $\lb^2$ modeled
by the integration of $\lb$ (eq. \ref{eq:modelRDT}) and the same quantity
directly computed from the resolved and 
sub-grid scales ($\lb=(\UU\cdot\nabla)
\uu^{\prime}+(\uu^{\prime} \cdot \nabla)\UU$).}
\label{fig:Lamb-M046-decaying}       
\end{figure}

One sees that the RDT model leads to a constant increase of 
the smallest modes of $\lb$
in time. After a given period of time, the contribution of these
unrealistic resolved scales of $\lb$ influence the model
of the velocity field resolved scales.\

An explanation of this feature can be found
following a recent study of the RDT model
(\ref{eq:SRDT}) \cite{dubrulle04a} showing that the
process of small-scale stretching by random large
scales is akin to a dynamo process, with
exponential increase of the small-scale energy. A
way to stabilize the system is to include a
friction term  in the RDT equation, leading to a
stationary energy spectrum with index depending
on the friction time $\tau_f$. A Kolmogorov
$k^{-5/3}$ spectrum is obtained for
$\tau_f=27/22 \, \Omega$, where $\Omega=\langle
\left(S_{ij} S_{ij}\right)^{-1/2} \rangle$ is a
typical stretching rate based on spatial average
of the large-scale velocity stress tensor
$S_{ij}$. Using eq. (\ref{eq:lamb}), one sees that such a friction term
generates an equivalent friction term in the
equation for the Lamb vector. This remark motivates the introduction
of a stabilizing friction term $-\lb/\tau_f$ in
the equation for $\lb$ to try and stabilize the
coupled system. 
Indeed one observes a significant
improvement with respect to the original RDT model. We therefore
adopt this procedure as our starting RDT model, from which we now
build our Langevin-LES model.

\subsection{Derivation of the  model} 

The derivation of the LES-Langevin model of turbulence proceeds in
two steps. In the first one, we replace the term $
\overline{(\uu^{\prime}\cdot\nabla)\uu^{\prime}}$
into a turbulent viscous term $\mu_t \Delta \UU$, acting only at large
scales, in the spirit of standard deterministic "eddy-viscosity"
models. In a second step, we derive a suitable Langevin equation of
the turbulent force $\lb$  through a decimation
of the number of degrees of freedom corresponding to scales beyond the
aliasing limit. For this, we introduce a strong hyperviscosity to
damp all components
beyond a given cut-off wavenumber $k_m$. There is a priori complete
freedom for the choice of $k_m$. Here, we note that the
cascade-driven forcing $\ff$ has components only up to $k=2k_c$.
Therefore, any component of $\lb$ beyond $3k_c$ will only be generated
through secondary processes (stretching) rather than through the
forcing. We may then hope that any $k_m$ between $2k_c$ and $3k_c$
provides the dominant contribution to the stochastic term $\lb$.
We found that $k_m=2k_c$ is in fact sufficient to capture
this dominant contribution. Our Langevin-LES (LRDT)  model is
therefore finally given by:

\begin{eqnarray}
\label{eq:LD-LES}
&&\partial_t \UU + \overline{( \UU\cdot \nabla) \UU}  +
\overline{\lb}= -\nabla P+\nabla(\nu+\mu_t)
\nabla \UU; \; [0<k<k_c] \nonumber\\
&&\partial_t \lb =-(\lb/\tau_f)-(\UU\cdot\nabla)
\lb^\prime -( \lb^\prime \cdot\nabla)\UU \nonumber \\
&&\quad \quad \;
+\nabla(\nu + \nu_t) \nabla
\lb+{\bf \xi_0}+{\bf \xi}; \; [0<k<2k_c] \\
&&{\bf \xi_0}=-(\UU\cdot\nabla)
   \ff-( \ff\cdot\nabla)\UU,\nonumber\\
&&<{\bf \xi}>={\bf 0},\quad
<\xi_i(t,x)\xi_j(t',x')>=F_{ij}(t-t',x-x'),\nonumber
\end{eqnarray}
where $\tau_f=27/22\langle \left(S_{ij}
S_{ij}\right)^{1/2} \rangle$, $\nu_t$ and
$\mu_t$ will be specified later and
$f_i=\partial_j\left(U_i U_j-\overline{U_i U_j}\right)$.
Looking at eq. (\ref{eq:LD-LES}), one recognizes a LES model where
the backscatter coming from resolved scales-subfilter interaction is
parameterized through a noise. The latter obeys a generalized Langevin
equation, with friction made of viscosity and rapid distortion by
resolved scales, and with stochastic forcing
${\bf \xi_0+\xi}$, generated through
the energy cascade and pressure processes.

\subsection{Approximations}
The RDT-based model we described is fully dynamics, and a priori
adapted to any type of geometry, with possible anisotropies. In
certain very simple cases, however, this model may support additional
approximations, enabling a reduction in the computation cost. We
present two examples below.

\subsubsection{Quasi-linear approximation (LQL)}
In the spirit of quasi-linear approximation, we may
lump the transport and stretching by resolved scales with the viscous
and friction term and simply replace
them by a total friction term $-\lb/\tau$, where $\tau$ is a typical time
scale to be chosen later.

Note that this simplification somehow relies on the hypothesis that
the advection and stretching of $\lb$ by the resolved scales
proceeds in an isotropic manner. We therefore do not expect this
simple procedure to be fully valid in cases with e.g. stratification,
or rotation, or walls. In the present case, it appears to give very
good results, provided the time scale is chosen appropriately. This
model therefore looks like:

\begin{eqnarray}
&&\partial_t \UU + \overline{(\UU\cdot \nabla) \UU}  +
\overline{\lb}= -\nabla P +\nabla(\nu+\mu_t) \nabla \UU; \; [0<k<k_c] \nonumber\\
&&\partial_t \lb =-(\lb/\tau)+{\bf \xi_0}+{\bf \xi}; \; [0<k<2k_c]
\label{eq:LS-LES}
\end{eqnarray}

\subsubsection{The over-damped approximation (LQLOD)}
In situation where the time scale $\tau$ is very small, the damping
in the Langevin equation is very large, and there is almost
instantaneous adjustment of $\lb$ to the forcing. This
situation typically arises in locations where the resolved
gradients are very large, i.e;. at the border of coherent resolved
eddies. It is therefore interesting to run the model in this
over-damped limit, to see whether these kinds of event dominate the
overall dynamics. The resulting model (Langevin
Quasi-Linear Over-Damped) looks like:
\begin{eqnarray}
&&\partial_t \UU + \overline{( \UU\cdot \nabla) \UU}  +
\overline{\lb}= -\nabla P+\nabla(\nu+\mu_t) \nabla \UU; \; [0<k<k_c] \nonumber\\
&& \lb =\tau\left({\bf \xi_0}+{\bf \xi}\right); \; [0<k<2k_c]
\label{eq:LSOD-LES}
\end{eqnarray}
Through this approximation, our model, based on the estimation of
$\lb$, becomes equivalent to the ADM velocity-estimation model of
Domaradzki and collaborators
\cite{domaradzki97,domaradzki99,domaradzki00}. In a recent improvement
of this model, the estimated small scales are used to integrate the
Navier-Stokes equation on the fine grid over a given period of time
short enough to prevent the pile up of the energy at the finest
scales. The ADM model can therefore be viewed as a special case of
our Langevin model. In the following, we show that this approximation
perform very well and is indistinguishable from the general model in
the isotropic homogeneous case. However, it may not be the case in
more general situations, such as near walls. We leave this for future
work.

\subsection{Parameters}
Our model includes three parameters $T_{ij}$, $\tau$ and $\mu_t$, that we now
discuss. As previously discussed, the stochastic
correlation $T_{ij}$ comes from pressure
contribution, and we found its influence be
irrelevant in the case with periodic boundary
condition. In the present study, we therefore
adopt $T_{ij}=0$, leaving its study for cases
involving walls. For the time scale, we set
$\tau=\tau_f+\alpha \left(S_{ij}
S_{ij}\right)^{-1/2}$, so as to allow spatial
variation of the friction time scale in agreement
with local stretching. Alternatively, one may
also use a global friction time by considering
only spatial average of the local stretching
rate, so that $\tau=\tau_m= \beta \tau_f$. This
last prescription is actually better in term of
computational cost. Empirically, we found that
both prescriptions lead to a good model. In our
numerical tests, we chose $\alpha=1$ or $\beta=1/2$ 
but we checked that the results are  not very sensitive
to these parameters

For the turbulent viscosity, we have an a priori rather vast choice,
ranging from spectral to Smagorinsky models. In the present case,
where our numerical scheme uses fast Fourier transform, the optimal
choice (in term of computational cost) is the spectral model.
The simplest choice is to link the turbulent viscosity to the kinetic energy
at the cut-off wavenumber $k_c$ ($\mu_t \propto \sqrt{E(k_c)/k_c}$). The
proportionality factor is tuned in our model and our numerical
tests. Since part of the dissipation is produced by $\lb$,
the proportionality constant turns out to be reduced
from 0.267 for the original spectral model
(see \cite{lesieur90}) to 0.08 in our case.
Summarizing, we get the following parameters for our models:

\begin{eqnarray}
&& T_{ij}= 0\\
\label{eq:parameter1} && \tau = (27/22)\langle \left( S_{ij} S_{ij}
\right)^{-1/2}\rangle+ \left( S_{ij} S_{ij} \right)^{-1/2} \\
\label{eq:parameter2} && \tau_m = 0.5 \langle \left( S_{ij} S_{ij}
\right)^{-1/2} \rangle\\
\label{eq:parameter3} && \mu_t = 0.08 \left(E(k_c)/k_c\right)^{1/2}
\end{eqnarray}

Models using the second prescription for the time scale will be
called Langevin Quasi Linear Averaged (LQLA). They are optimal in
term of computational cost.

\section{Model performances}
\label{sec:VN}
\subsection{Comparison of the Langevin models}
As a preliminary assessment of the interest of our model, we choose
to test it in a simple configuration where we have both the code and
the competence to implement it over a short time scale. This case
corresponds to homogeneous, isotropic turbulence with periodic
boundary conditions. It is clearly a restricted problem, where
backscatter has proved to be secondary, and where eddy-viscosity
based approaches perform very satisfactorily. In some
sense, this ideal case is therefore a challenge to our model, for it
is not clear whether there is room for further improvement with
respect to eddy-viscosity description. In the following, we
demonstrate a slight but discernible positive effect of our strategy.\

The test is done both on decaying turbulence and forced
turbulence in order to check the model ability to accurately
reproduce the energy cascade from resolved to unresolved scales.
The first comparison between the LRDT model (eq. \ref{eq:LD-LES}) and
the DNS is conducted over the energy spectra in  fig.
\ref{fig:M061-decaying} and the decay of the total kinetic energy  in
fig. \ref{fig:M061-decaying-nrj}.
\begin{figure}
\centerline{\includegraphics[width=0.75\columnwidth]{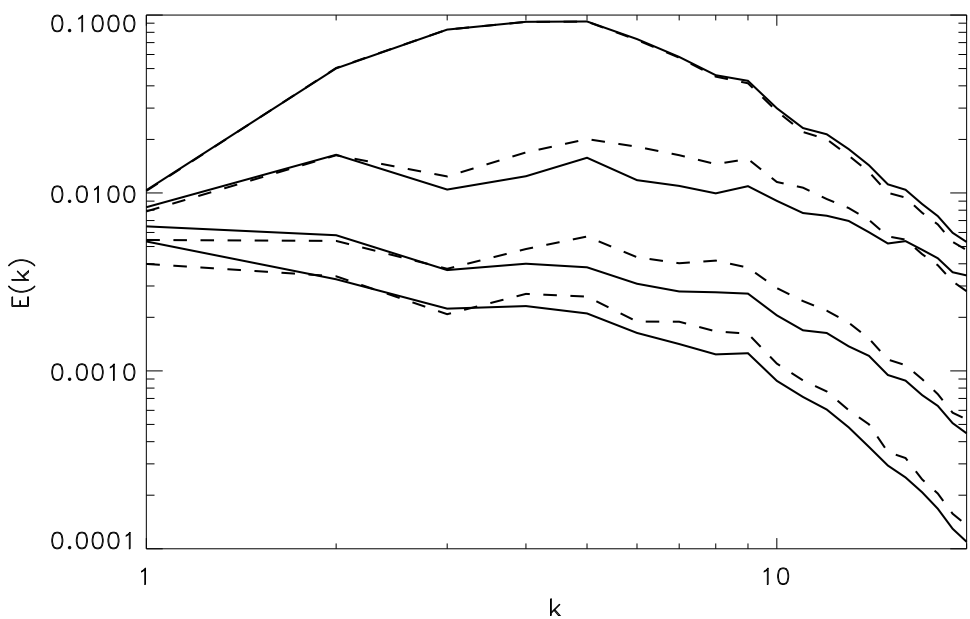}}
\caption{Energy spectra of the LRDT model (dash lines) and the equivalent 
DNS (continuous lines) of a decaying homogeneous isotropic turbulence.
The spectra are  compared for 4 different times: $t=(1,3,6,9)$.}
\label{fig:M061-decaying}       
\end{figure}
\begin{figure}
\centerline{\includegraphics[width=0.75\columnwidth]{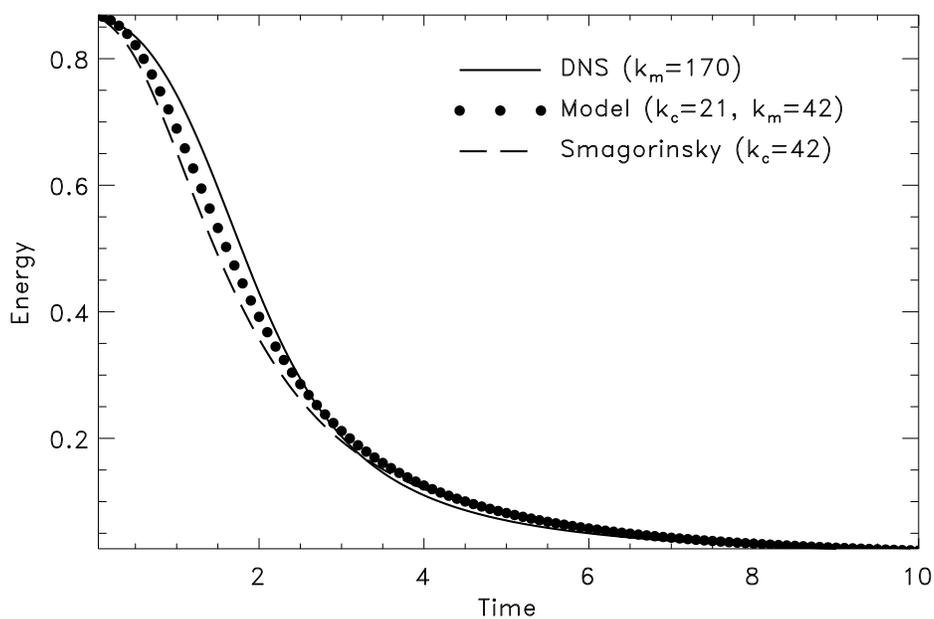}}
\caption{Comparison of the decay of
kinetic energy between the DNS of decaying turbulence, the LRDT
model (with $k_c=21, k_m=2 k_c$) and the Smagorinsky model with
$k_c=42$. The time $t=10$ correspond to a non-dimensional time of 6.3}
\label{fig:M061-decaying-nrj}       
\end{figure}
In this comparison, the resolved scale turbulent viscosity $\mu_t$ is
tuned to give the best result in terms
of energy spectra and decay of kinetic energy. The decimation of the
small-scale of $\lb$ (so as to keep only $2k_c$ modes in
each direction) is done through an hyper-viscosity ($\nu_t= 0.002 \, k^2
\sqrt{E(k_c)/k_c}$) super-seeding the turbulent viscosity of the
original RDT model. The energy
spectra are in good agreement with the DNS for $t < 1$ and slightly
differ for the smallest resolved scales at later time. We are
confident that a more elaborate tuning of the viscosity and friction term
could have improved the model over
later time, but such accuracy was not needed in the sequel.\\

The same numerical tests have been performed for the approximate
models, LQL and LQLOD, (figs. \ref{fig:M051-decaying} and
\ref{fig:M052-decaying}), with the same definition of the time
scale $\tau$ (eq. (\ref{eq:parameter1}),(\ref{eq:parameter2})). The results
are slightly less accurate than the original LRDT model, but the
two approximated models represent a good compromise between accuracy
and CPU cost. As discussed in the previous section, an alternative
model to LQL can also be formulate using an averaged value  $\tau_m$
of the time scale $\tau$. A comparison of averaged energy spectra between
LQL and LQLA models and DNS is done fig. \ref{fig:T051-T055-DNS-forced}
for a simulation of forced homogeneous isotropic turbulence. In the
simulation with LQLA model, the averaged value of the time scale $\tau_m$
is of the order of 0.1 turnover time. The two LES simulations are very 
similar and are both in good agreement with the DNS.
\begin{figure}
\centerline{\includegraphics[width=0.75\columnwidth]{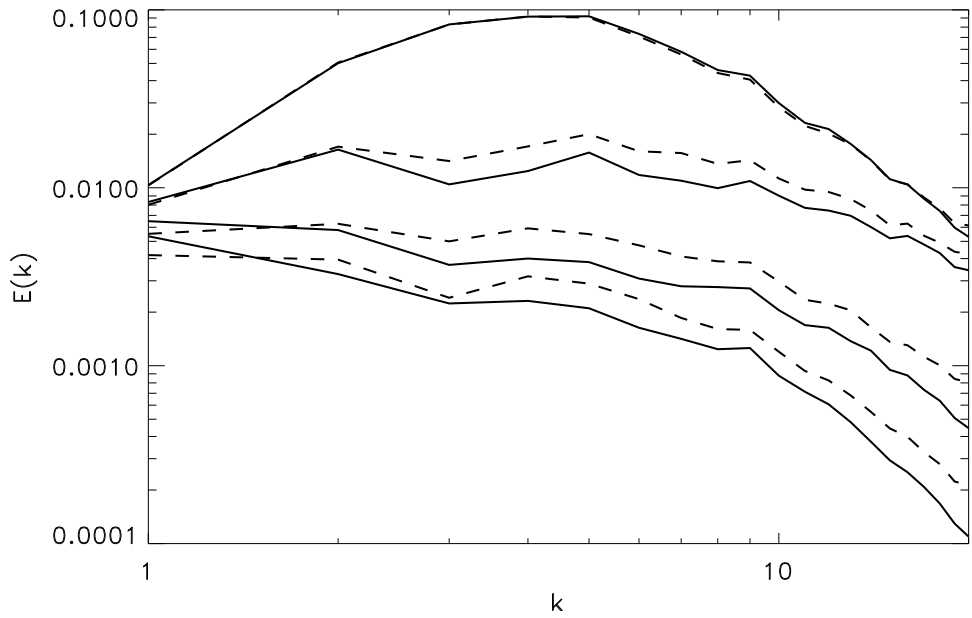}}
\caption{Energy spectra of the LQL model (dash lines) and the
 equivalent DNS (continuous lines) of a decaying
homogeneous isotropic turbulence. The spectra are 
compared for 4 different times: $t=(1,3,6,9)$. }
\label{fig:M051-decaying}       
\end{figure}
\begin{figure}
\centerline{\includegraphics[width=0.75\columnwidth]{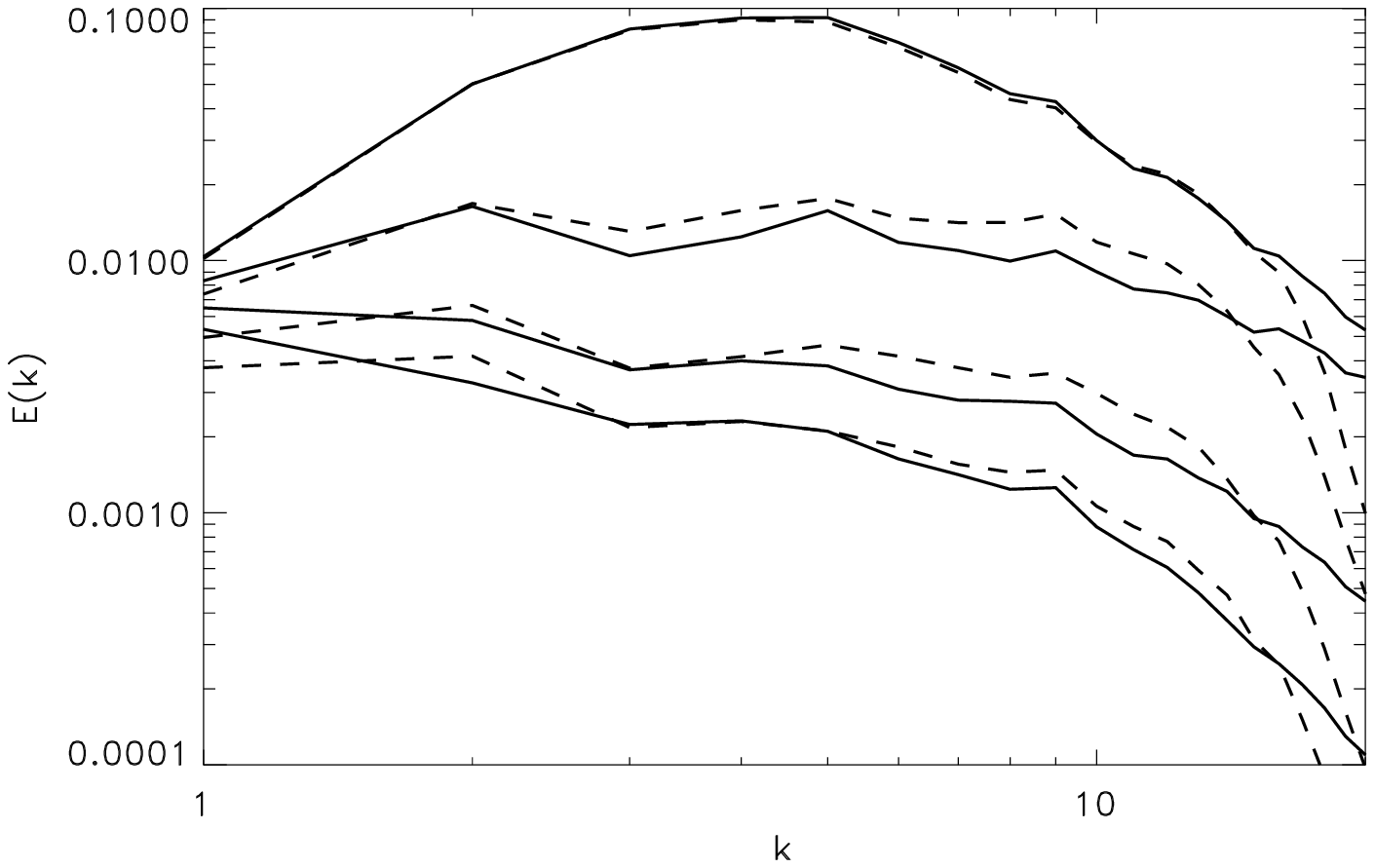}}
\caption{Energy spectra of the LQLOD model (dash lines)
and the equivalent DNS (continuous lines) of a decaying homogeneous isotropic
turbulence. The spectra are compared for 4 different times: $t=(1,3,6,9)$.}
\label{fig:M052-decaying}       
\end{figure}
\begin{figure}
\centerline{\includegraphics[width=0.75\columnwidth]{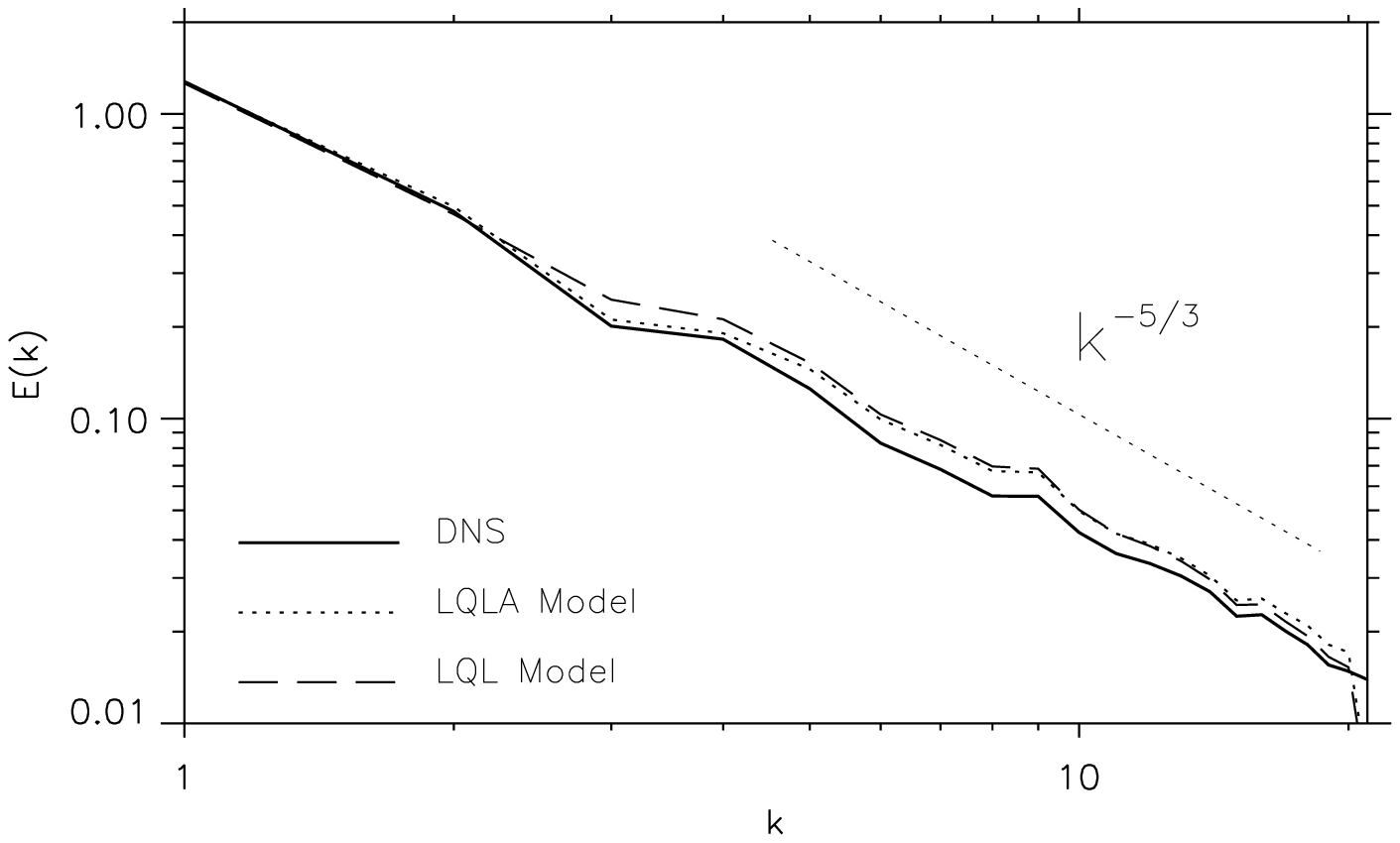}}
\caption{Averaged energy spectra for
the simulation of forced isotropic homogeneous turbulence.
The spectra are averaged over 4 turnover times for the DNS
and 40 turnover times for the three LES. The DNS
was performed with $384^3$ Fourier modes.}
\label{fig:T051-T055-DNS-forced}       
\end{figure}
\subsection{Energy transfer}
One of the ability of the model compared to eddy viscosity model is to allow
energy transfer from subfilter scales to resolved scales through the
model of the cross stress tensor by the Langevin equation of $\lb$. The
amount of backscatter can be quantify  by:
\begin{equation}
T_{res,sub}(\kk) = \int \int \UU(\pp)
\lb_{\perp}(\qq) \delta(\kk-\pp-\qq) d^3\pp d^3\qq
\label{eq:transfer}
\end{equation}
The integration of $T_{res,sub}(\kk)$ over a
shell  leads to the energy transfer
as a function of the $k$. This quantity was investigated for the LQLA model
in the case of forced isotropic turbulence (see
fig. \ref{fig:transfer}).
\begin{figure}
\centerline{\includegraphics[width=0.75\columnwidth]{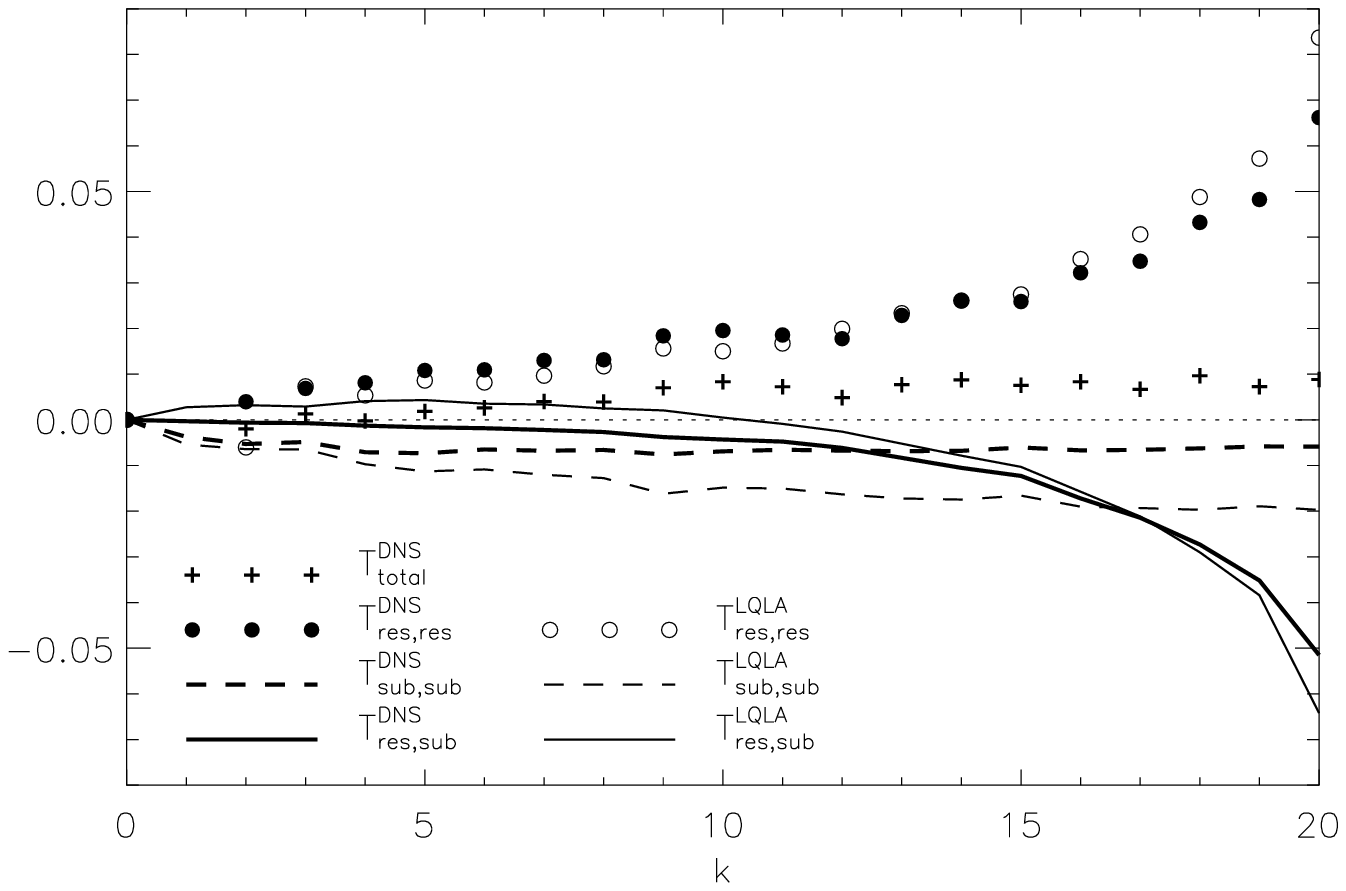}}
\caption{Comparison of the energy tranfer between
the DNS and the LQLA model for forced isotropic turbulence. The energy
tranfer is split into the 3 components involving resolved and subgrid
scales. $T^{LQLA}_{res,sub}$ is the energy transfer from the Lamb vector
(see eq. \ref{eq:transfer}) and $T^{LQLA}_{sub,sub}$ is the disipation
introduced by the turbulent viscosity $\mu_t$.}
\label{fig:transfer}       
\end{figure}
The intensity of the energy transfer through the Lamb vector
$T^{LQLA}_{res,sub}(\kk)$ is compared with the
energy transfer from resolved scales
$T^{LQLA}_{res,res}(\kk)$ and the energy
dissipation introduced by the model of the Reynolds stress tensor with
a turbulent viscosity $T^{LQLA}_{sub,sub}(\kk)$.
Each component of the energy transfer is also
compared with the equivalent component computed
from the DNS. The statistics have been conducted over 4
turnover times after stabilization of the
statistics. The corresponding average energy spectra of the
two  simulations are shown in Fig.
\ref{fig:F512-T055-TLES-T053-forced}.
\begin{figure}
\centerline{\includegraphics[width=0.75\columnwidth]{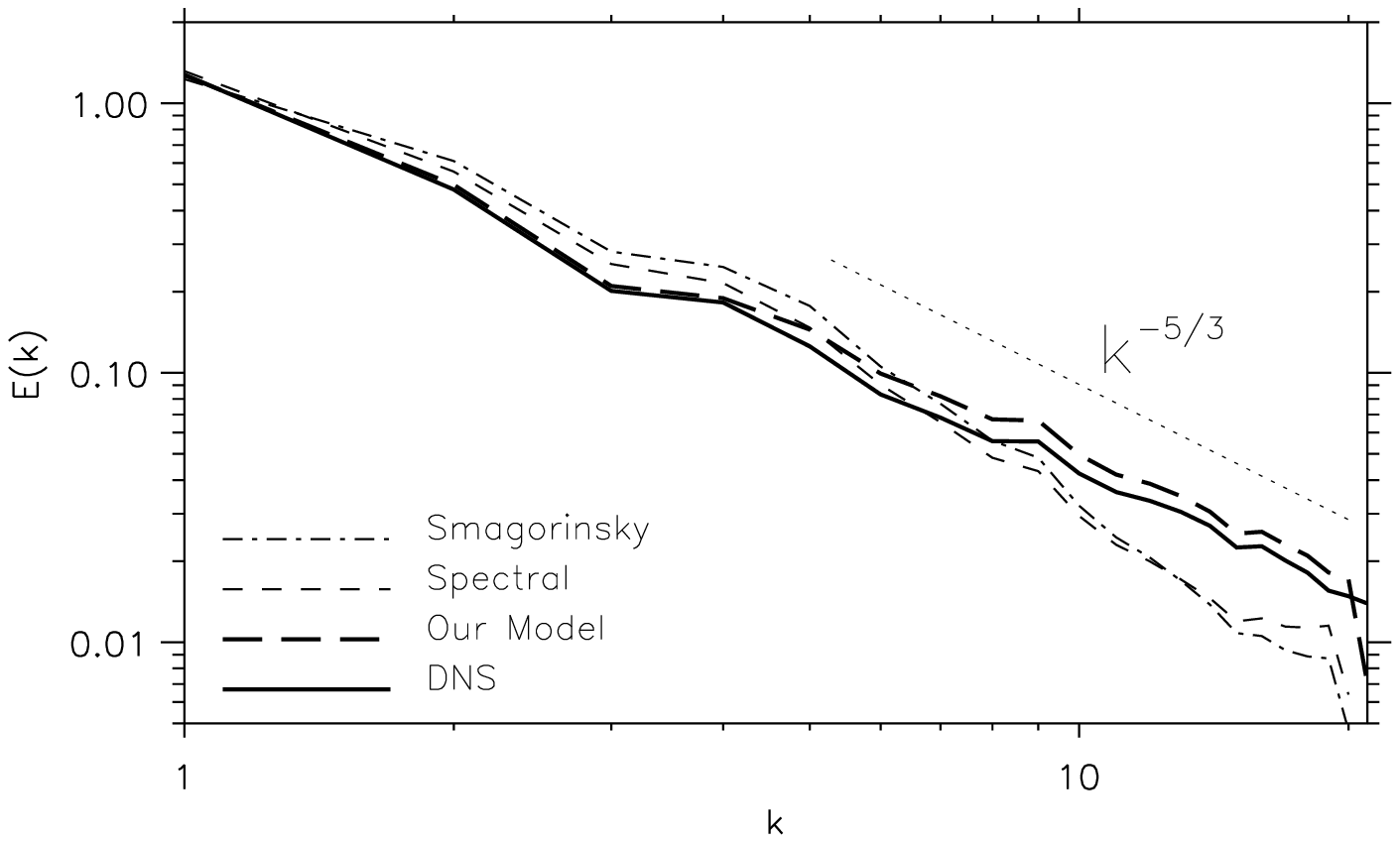}}
\caption{ Comparison of the energy spectra of our
LQLA model with the DNS, the Smagorinsky model and the
spectral model (averaged over 4 turnover times for DNS
and  40 turnover times for the 3 models)}
\label{fig:F512-T055-TLES-T053-forced}       
\end{figure}
Since the two solutions used for the statistics
differ in time, it is difficult to compare
exactly the transfer between the model and the DNS. However one can
compare the sign and the order of magnitude of
each component. The transfer coming from the Lamb vector exhibits
a small amount of backscatter, which is not
present in the DNS. However the transfer is very comparable at higher
wavenumbers. The turbulent viscosity $\mu_t$ used
to model the Reynolds stress tensor leads to a dissipation of
energy higher than the equivalent energy transfer
from the DNS. The model of this term could probably be improved.

\subsection{Comparison with other LES models}
Within the present framework (isotropic, homogeneous turbulence)
it therefore seems that the LQLA model represents the best compromise
in terms of accuracy and CPU cost. We now explore its performances
with respect to other classical LES models in this field.

The simpler LES models for homogeneous, isotropic turbulence are
based on eddy-viscosity concept, namely the Smagorinsky model
\cite{smagorinsky63} and the spectral model
\cite{lesieur90}. The later has only been tested
for forced isotropic turbulence. The former corresponds to a
Smagorinsky model (without dynamical procedure) in which the
constant was tuned to get the best results
in terms of energy decay. It was run on the finest
grid of our model (the grid used for the integration of $\lb$). The
comparison is done using several statistical diagnostic, tracing
different properties of the model.

\subsubsection{Energy spectra}

A comparison of the energy spectra
for these two models and our LQLA model is shown  in
fig. \ref{fig:M055-decaying} for the decaying case and in
fig. \ref{fig:F512-T055-TLES-T053-forced} for the forced case.
\begin{figure}
\centerline{\includegraphics[width=0.75\columnwidth]{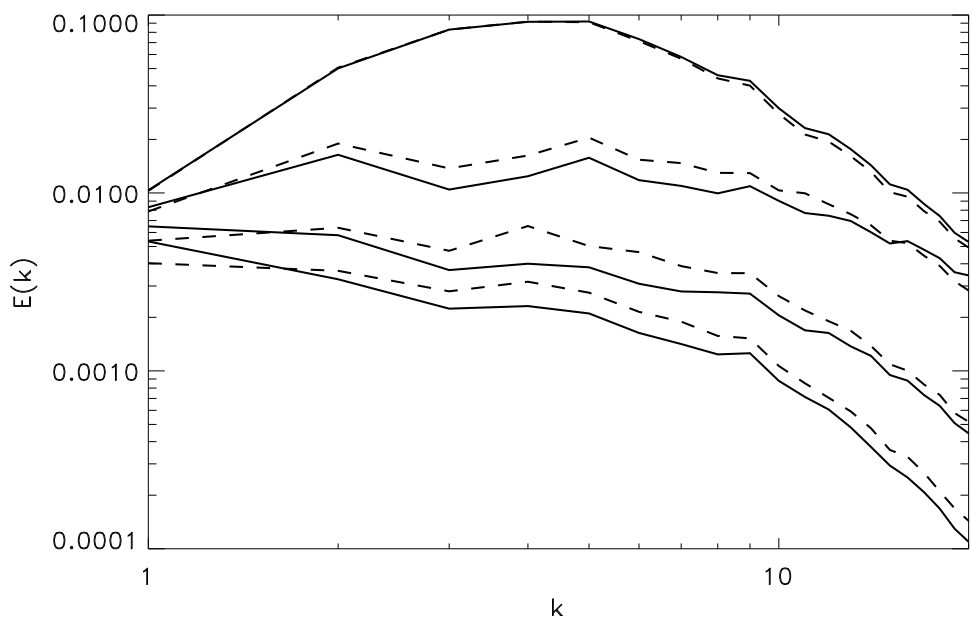}}
\caption{Energy spectra of the LQLA model (dash lines) and the
 equivalent DNS (continuous lines) of a decaying homogeneous isotropic
turbulence. The spectra are compared for 4 different times: $t=(1,3,6,9)$.}
\label{fig:M055-decaying}       
\end{figure}
In the decaying case, our Langevin model
performed at least as well as the Smagorinsky model in terms of
energy decay. In the forced case, the
Smagorinsky model clearly overestimate the energy dissipation at the
smallest resolved scales, resulting in an energy spectrum much steeper than the
theoretical $-5/3$ slopes. Similar result is found for the spectral
model, while the LQLA model compares much better with the
DNS at all scales. Indeed, our model seems to model properly
the energy transfer at the largest scales and the
energy dissipation near the cut-off.

\subsubsection{Skewness and flatness}

We also compared the skewness and the flatness of the LES models
to the same statistics for the corresponding scales of the DNS
in the case of forced turbulence. The results are displayed in
table \ref{tab:stat}. The LQLA model gives a correct sign of the
skewness, with a value three times too large, and a correct estimate
of the flatness. By comparison, the Smagorinsky model predicts a
wrong sign for the skewness and underestimates slightly the flatness.
The spectral model gives the best estimate of the skewness (with
correct sign) and underestimates the flatness of the distribution.

\begin{table}[h]
\begin{center}
\caption{\label{tab:stat}  Skewness and Flatness of the resolved
scales of the LQLA model, the spectral model, the Smagorinsky model
and the corresponding  DNS in the case of forced turbulence.}  
\vspace{1cm}
\begin{tabular}{|l|c|c|}
\hline
model       & skewness  & flatness \\
\hline
DNS         & -0.003314 & 0.3141 \\
LQLA        & -0.011198 & 0.3107 \\
Spectral    & -0.004328 & 0.3094 \\
Smagorinsky &  0.013744 & 0.3078 \\
\hline
\end{tabular}
\end{center}
\end{table}

\subsubsection{Probability distribution function}

    A more refined statistical comparison can be
performed using the Probability
Density Function of the velocity increment ${\delta v}_r =
(\vv(\xx+\rr) + \vv(\xx))$ (figs. \ref{fig:pdfl} and
\ref{fig:pdft}).
\begin{figure}
\centerline{\includegraphics[width=0.75\columnwidth]{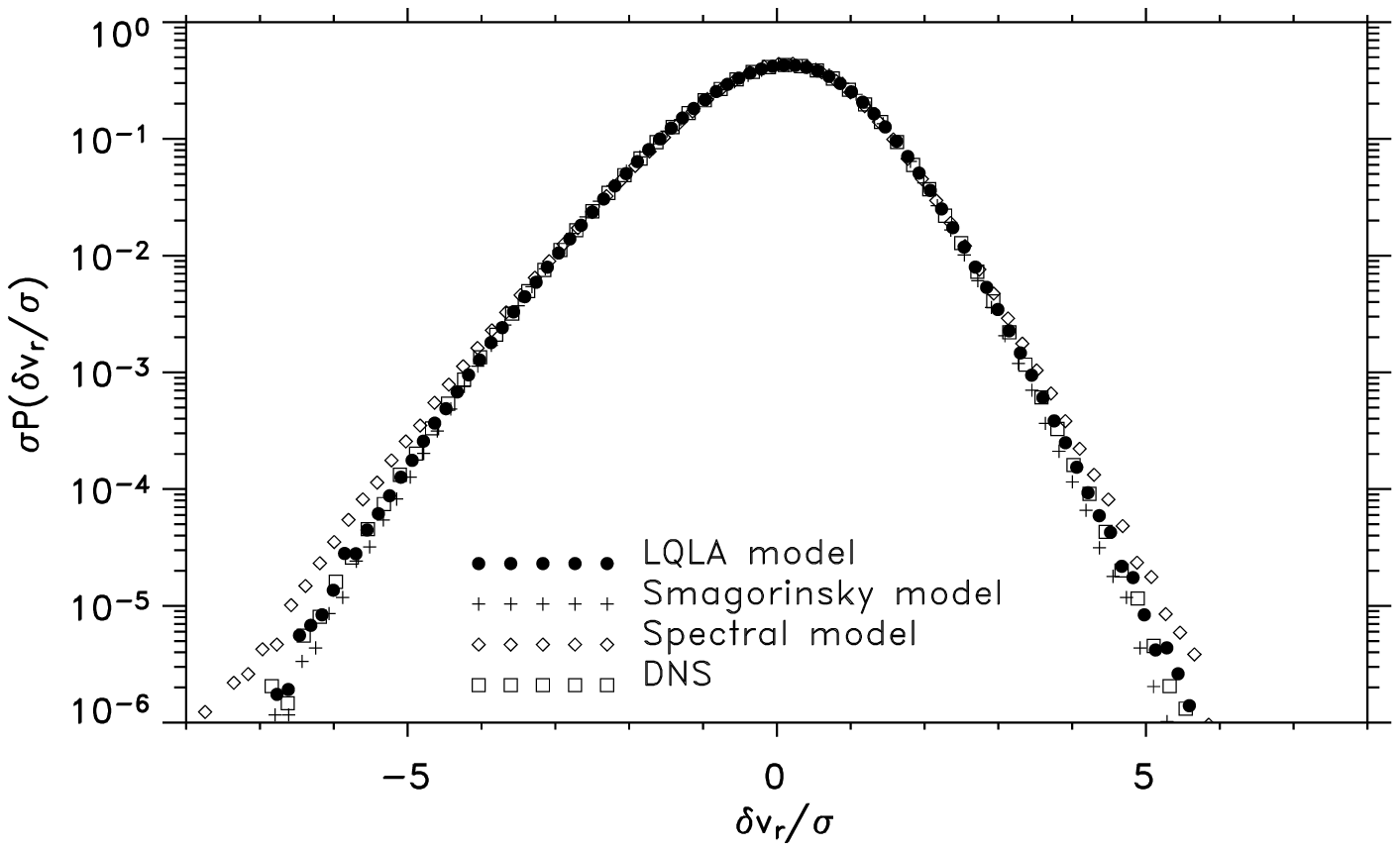}}
\caption{ Comparison of the normalized PDF of longitudinal 
velocity increments ($\delta v_r = \vv(\xx+\rr) - \vv(\xx)$ with
$\vert\rr\vert=\frac{L}{64}$ and $\rr\times\vv=0$) for our LQLA
model, the Smagorinsky model, the spectral model and the DNS. The PDF
are averaged over  4 turnover times for DNS and 40 turnover times for
the LES models.}
\label{fig:pdfl}       
\end{figure}
\begin{figure}
\centerline{\includegraphics[width=0.75\columnwidth]{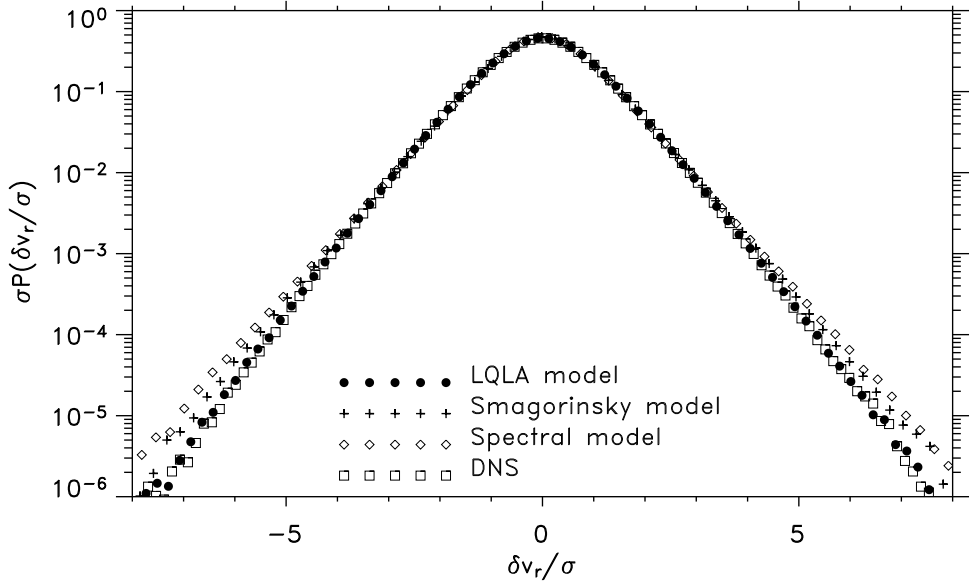}}
\caption{Comparison of the normalized  PDF of transverse velocity
increments ($\delta v_r = \vv(\xx+\rr) - \vv(\xx)$ with
$\vert\rr\vert=\frac{L}{64}$ and $\rr\cdot\vv=0$) for our LQLA model,
the Smagorinsky model, the spectral model and the DNS. The PDF are
averaged over  4 turnover times for DNS and 40 turnover times for the
LES models.}
\label{fig:pdft}       
\end{figure}
Because of the coarse resolution of the LES
simulations and the small number of statistics
for the equivalent scales of DNS, the
differences of PDFs between the models and the DNS are difficult to
observe. However, the statistics with the LQLA
model seems to be in a closer agreement with the
DNS for both longitudinal and transverse
velocity increment than the two other dissipative models.

\subsubsection{Statistics of invariants}

A different statistical test can also be performed on
the tensor of velocity gradient tensor $A_{ij}= \partial \overline{u_i} /
\partial x_j$. We choose to focus on the two following invariants:
\begin{eqnarray}
Q &=& - \frac{1}{2} \overline{A}_{im} \overline{A}_{mi} \\
R &=&  - \frac{1}{3} \overline{A}_{im} \overline{A}_{mk} \overline{A}_{ki},
\end{eqnarray}
which have been used to discriminate between different turbulent
models \cite{vanderbos02}. For this, we normalize the two invariants
with a time scale based on the strain-rate tensor.
The two invariants become:
\begin{eqnarray}
Q^* &=& \frac{Q}{\langle \overline{s}_{ij} \overline{s}_{ij} \rangle} \\
R^* &=& \frac{R}{\langle  \overline{s}_{ij} \overline{s}_{ij} \rangle^{3/2}}
\label{eq:invariant}
\end{eqnarray}
where $\overline{s}_{ij} = \overline{S}_{ij} - \langle
\overline{S}_{ij} \rangle$
and $\langle \cdot \rangle$ is a sample averaging.
The figure \ref{fig:rq} shows the joint PDF of $Q^*$ and $R^*$ for the resolved
scale of the LQLA model, the Smagorinsky model,
the spectral model, and the DNS. One sees that
the LQA model and the spectral model
reproduce better the joint PDF than the Smagorinsky model, which
tends to squeeze the level lines towards $R=0$, a feature already
noted in \cite{vanderbos02}.
\begin{figure}
\begin{minipage}{0.49\columnwidth}
\includegraphics[width=\columnwidth]{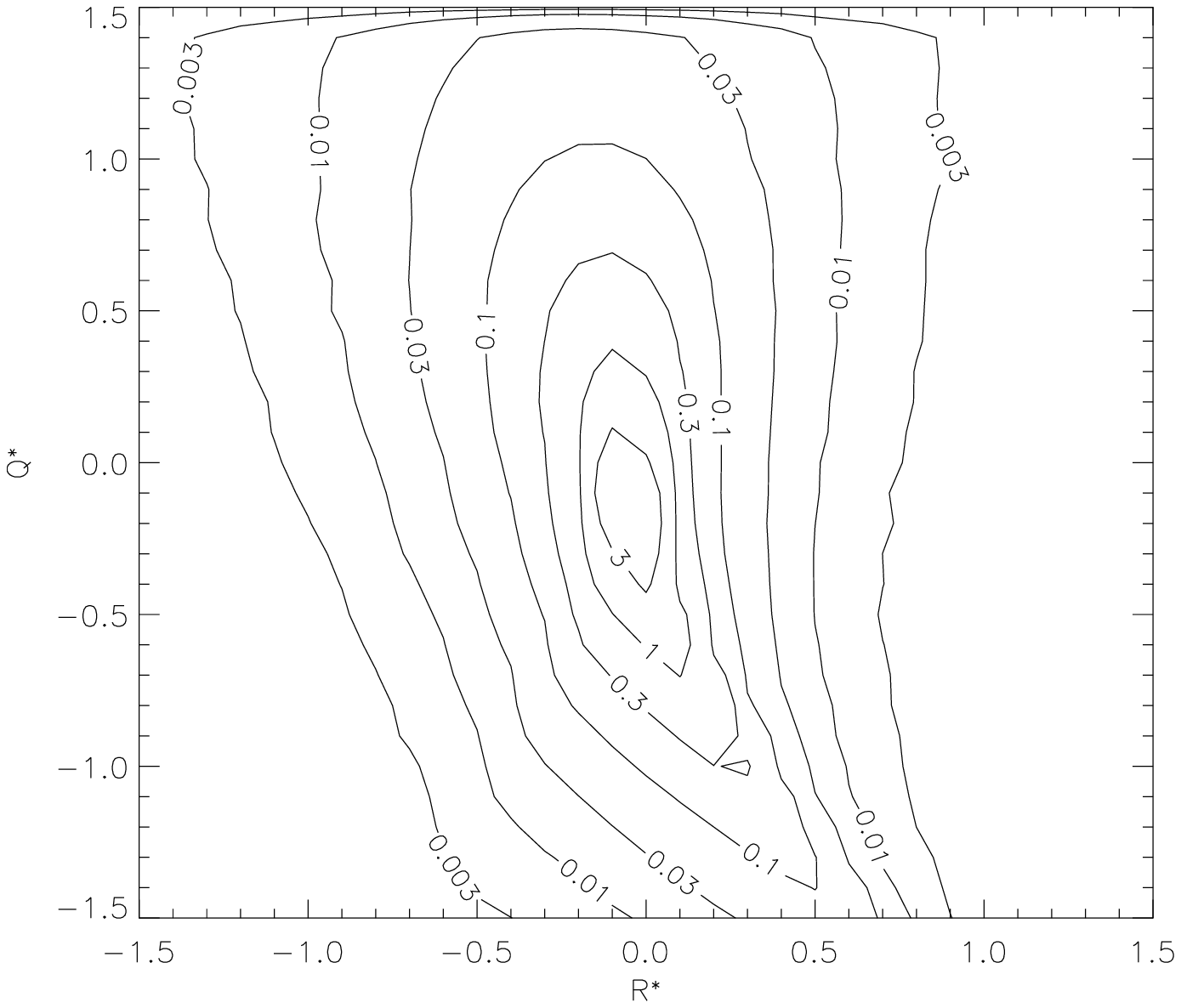}
\end{minipage}
\hfill
\begin{minipage}{0.49\columnwidth}
\includegraphics[width=\columnwidth]{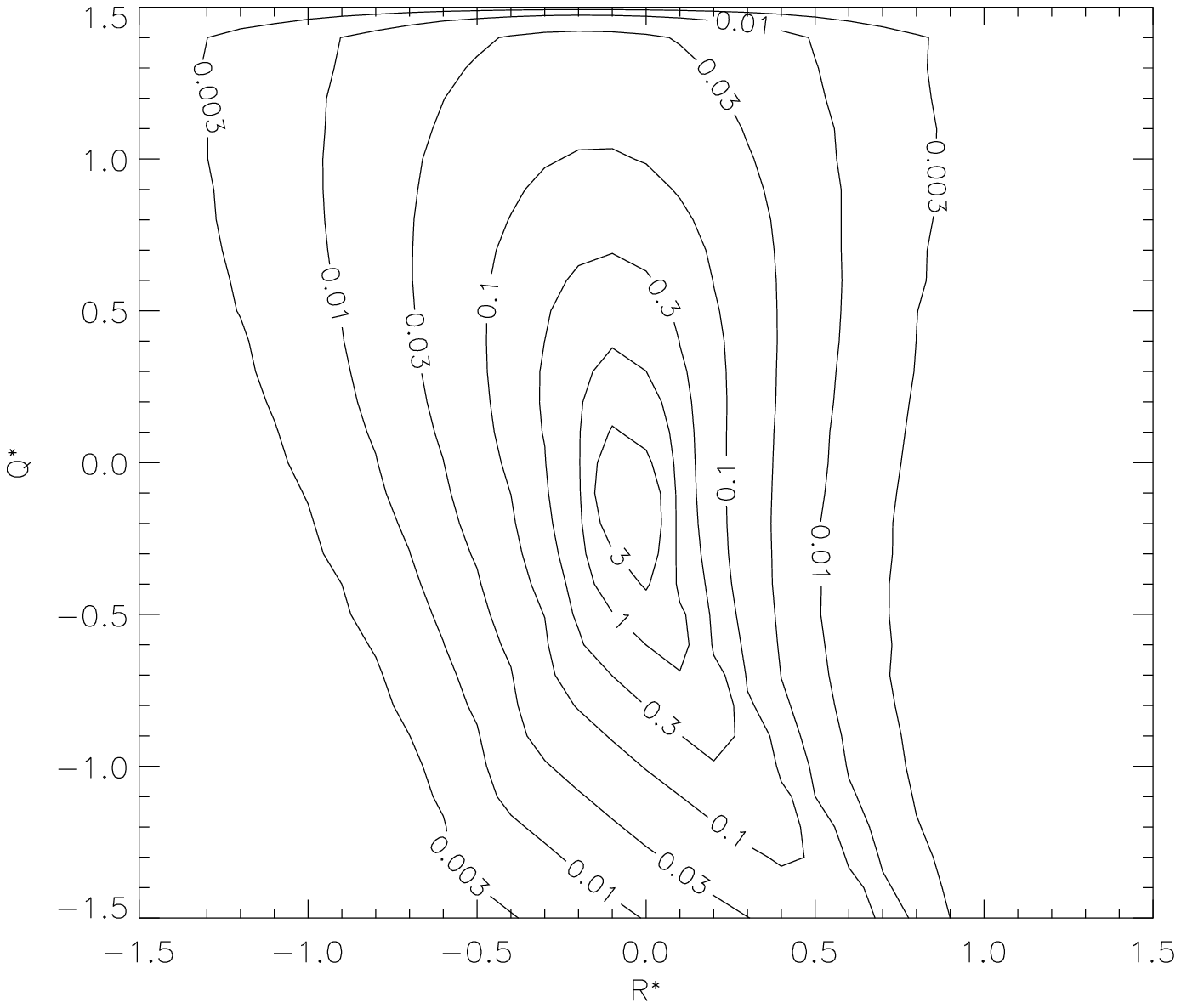}
\end{minipage}
\begin{minipage}{0.49\columnwidth}
\includegraphics[width=\columnwidth]{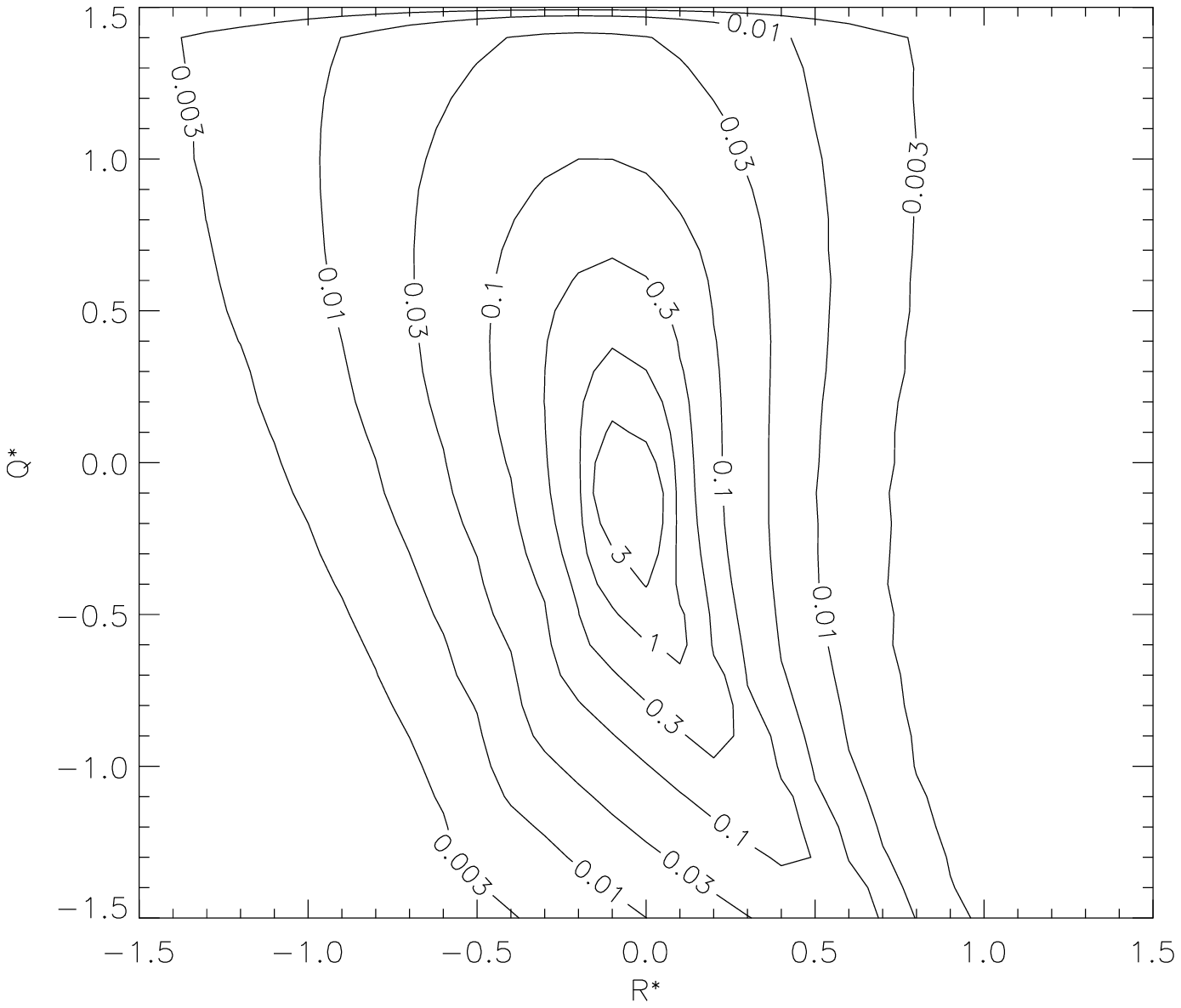}
\end{minipage}
\hfill
\begin{minipage}{0.49\columnwidth}
\includegraphics[width=\columnwidth]{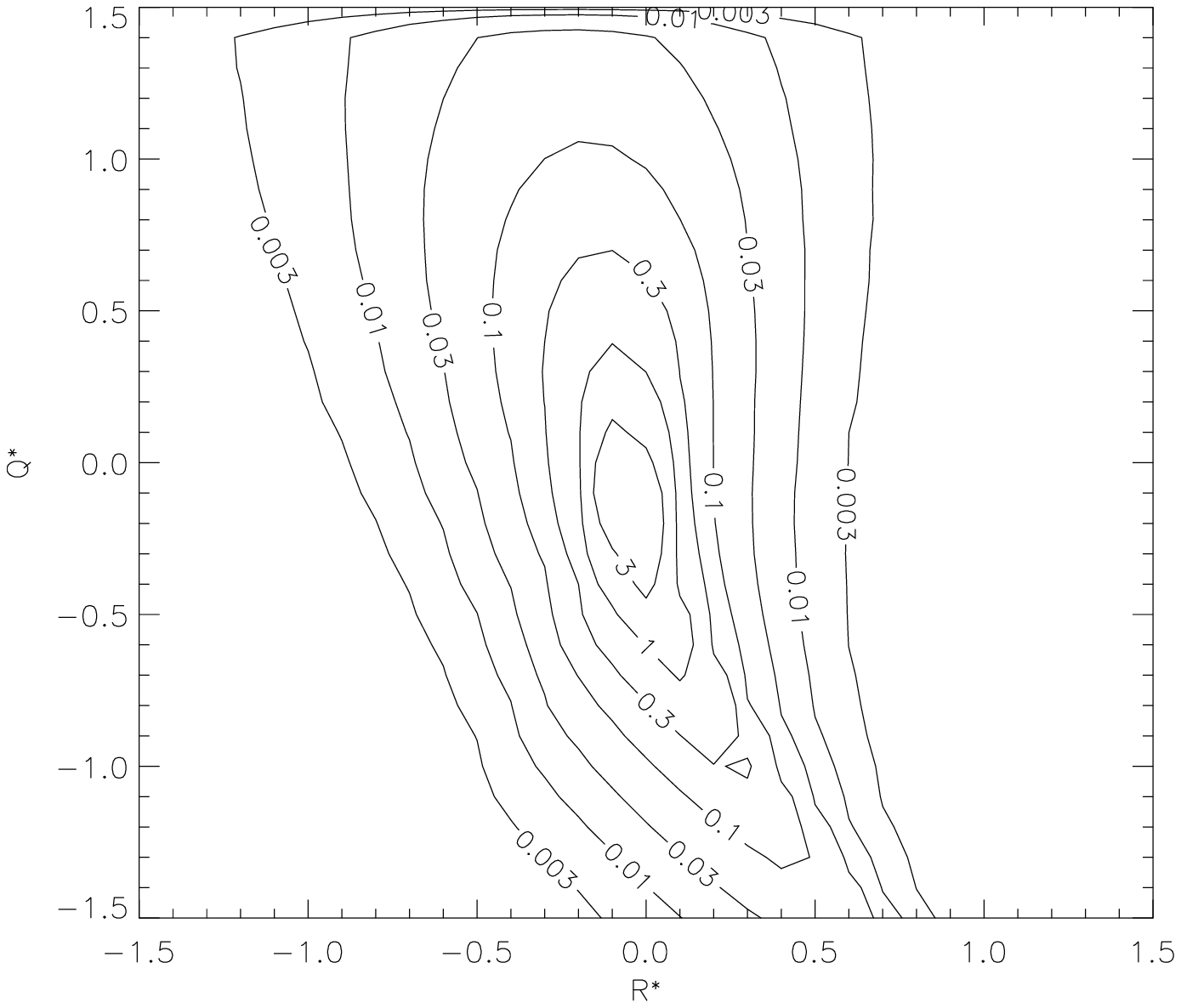}
\end{minipage}
\caption{Comparison of the joint PDF of $R^*$ and $Q^*$
(eqs. \ref{eq:invariant}) for the resolved
scales of the LQLA model (upper right), the
Smagorinsky model (lower right), the spectral
model (lower left) and the corresponding DNS
(upper left) in the case of  forced turbulence.}
\label{fig:rq} 
\end{figure}

\subsubsection{Strain statistics}

Another important characteristic of the strain rate tensor S is the statistics
of its eigenvalues. To quantify it, Lund and Rogers \cite{lund94}
proposed the following parameter:
\begin{equation}
s^* = -3 \sqrt{6} \frac{\alpha \beta \gamma}{(\alpha^2 \beta^2 \gamma^2)^{3/2}}
\label{eq:s-star}
\end{equation}
where $\alpha$, $\beta$ and $\gamma$ are the eigenvalues of $S$ such as
$\alpha< \beta< \gamma$. The term $s^*$ measure the shape of the deformation
caused by the strain rate tensor, with $s^*=-1,0,1$ corresponding
respectively to worms, shear and pancakes. The probability density
function of $s^*$
for the resolved scales of the LQLA model as well as for the Smagorinsky and
the spectral models are computed in the case of forced turbulence.
The comparison is shown on figure
\ref{fig:s-star}. The comparison shows very good agreement for both
the LQLA model and the spectral model with respect to the DNS, while
the Smagorinsky model shows poorer agreement. \\
\begin{figure}
\centerline{\includegraphics[width=0.75\columnwidth]{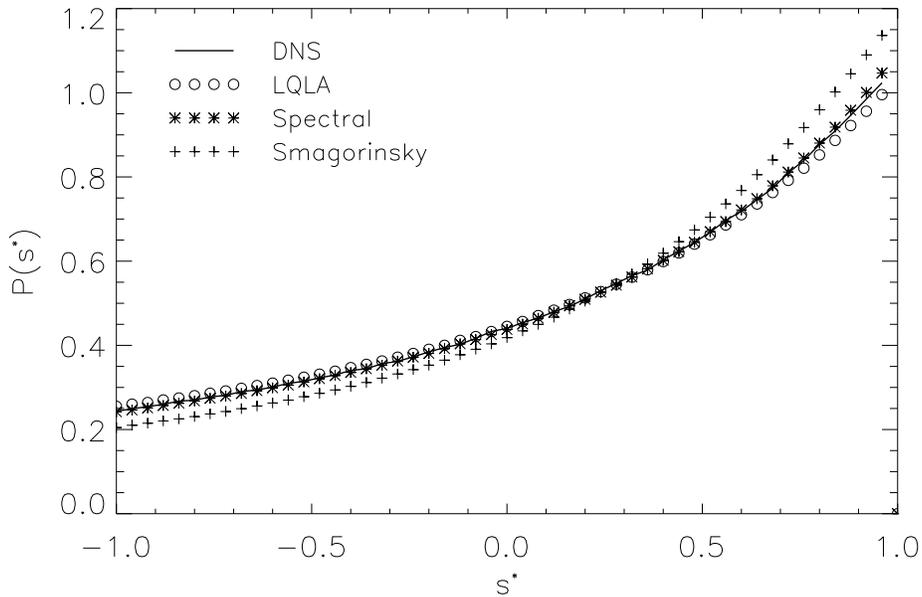}}
\caption{Comparison of the  PDF of $s^*$
(eq. \ref{eq:s-star}) for the resolved scales of the LQLA model,
the Smagorinsky model and the spectral model and the corresponding
DNS in the case of forced turbulence.}
\label{fig:s-star}       
\end{figure}

\subsubsection{Computational cost}
The performance of each model can also be analyzed with respect to its
computational cost. However, we are not able to give precise estimations
of the performance because the implementation of each model was not
optimized in terms of CPU cost. Moreover, the cost is highly related to the
type of discretization and the numerical scheme.
But, looking at the equations,
we can expect an optimal computational cost for our model to be of
the order of a DNS on a grid corresponding to the grid used for the integration of
$\lb$. It as been shown that an integration with wavenumbers up to $2 k_c$ is enough
for the equation of $\lb$. With the over-damped version of the model,
there is no integration for $\lb$ and we could
expect a gain of a factor 2 relative to the time step.

\section{Discussion}
\label{sec:CONCL}
In this work, we present a new strategy to model subfilter scale in a
Large-Eddy-Simulation. Our philosophy derives from the Brownian
motion, where the extra degrees of freedom are modeled through the
combination of a deterministic friction and stochastic forcing. Our
model therefore includes both a deterministic eddy-viscosity and a
stochastic turbulent force, solution of a generalized Langevin
equation. Through this combination, we aim at reproducing both the
transport enhancement through turbulent motions, and stochastic energy
backscatter from subfilter to resolved scale. Several model have been
proposed in the past, to reproduce these two effects. At this point,
it is interesting to review briefly the most popular ones, so as to
understand their limitations and differences with respect to our model.
A first example is the similarity model, which does
not assume the alignment of the turbulent stress tensor with the
strain rate tensor. In the original model of Bardina
\cite{bardina80}, the model of the Reynolds stress is achieved through
approximation of the unfiltered velocity by the
filtered velocity. This model can
be seen as the zero-level of more general deconvolution models that
rebuilds the unfiltered velocity (see \cite{germano86} for
a discussion on de-filtering in LES). This procedure however is also
subject to serious limitations, since a LES with an exact
deconvolution of the velocity field would
lead to a simulation of the Navier-Stokes equation on a coarse grid
without any model but with a sharp filter. Practically, the resulting
lack of dissipation is provided through the physical or the
numerical approximations of the deconvolution \cite{stolz99}. Other
concepts arose from a detailed analysis of the
energy transfer between resolved and subfilter scales. This leads to the
conclusion that the scales smaller than half the smallest resolved
scales have only few impact on the direct energy transfer to the
largest scales \cite{domaradzki87,zhou93,domaradzki94}. Following
this observation, a new method to capture the backscatter has been
proposed through direct estimation or a more accurate integration of the
largest subfilter-scales velocities. The simulations with these models
can be accurate but expensive LES or cheap approximated DNS. The dynamics
multilevel model \cite{dubois99,temam96,dubois98b} is an example of
cheap DNS as it requires the same resolution as the corresponding
DNS. Our 2D version of the RDT model \cite{laval04a} using a
Lagrangian evolution of small-scale wave-packets is more flexible since
the accuracy of the model is function of the number of modes
used for the subfilter scales. Better performances in terms of
computational cost can be obtained in models where only the largest
unresolved scales are estimated. An example is provided by the ADM
model of Domaradzki
\cite{domaradzki97,domaradzki99,domaradzki00}, where subfilter scales
are estimated on a finer grid. The estimation of subfilter scales is
performed after a first step of deconvolution. In a recent improvement
of this model the estimated small scales are used to integrate the
Navier Stokes equation on the fine grid over a given period of time short enough to
prevent the pile up of the energy at the finest scales. Finally, one
may note that attempt of direct inclusion of the backscatter have
been done in the past through empirical addition of random numbers,
with well chosen spectrum \cite{leith90,mason92}.  These models led
to noticeable improvement with respect to eddy-viscosity type of
approach for boundary layer or plane shear mixing layer.\

Most of these models have been validated and
compared in several flow configurations.
The most intensive comparison have been performed for isotropic turbulence
where the DNS can reach higher Reynolds number. For instance, Fureby et al
\cite{fureby97} made a comparative study of subgrid
scale model of eddy-viscosity type, models of
similarity type
and Miles model. For moderate Reynolds
number, they concluded that the difference between
LES with different models are small but not 
insignificant. The characteristics
of more specific type of models like eddy-viscosity models or similarity models
have been studied intensively by many authors
(\cite{liu94,meneveau99,lesieur96b}

From this review, it is clear that our originality is to seek for a
systematic derivation of the stochastic effect, rooted in the
Navier-Stokes dynamics.
By this mean, we hope both to avoid uncontrolled, empirical
modeling, and respect of all the symmetries of the original
Navier-Stokes equation (a constraint not always easy to respect, see
\cite{oberlack01}). Our model may also easily be generalized to other more
complicated systems, like e.g. rotating, stratified or inhomogeneous turbulence.
Specifically, we separate the subfilter scale
contribution into a term that correlates
well to resolved strain tensors, and a term susceptible to the
backscatter. The first term is modeled through a traditional
eddy-viscosity, while the second is modeled through a noise, obeying a
generalized Langevin equation. This equation is not postulated, but
derived from the Navier-Stokes equation, through hypothesis akin to
rapid distortion theory. The stochastic retro-action couples the
resolved scales and the noise through the energy cascade, ensuring strong
non-linearity of our model and non-trivial equilibrium solutions. A
corner stone of our model is the estimation of $\lb$, rather than
the velocity, like in ADM model. Apart from this difference, there is
some similarity between our model and the ADM model, in the simplest
approximations we considered (quasi-linear and over-damped
approximation). However, our model derives from the original
Navier-Stokes equation through a well-defined and systematic
procedure. \

     We have proposed a preliminary test of our model in isotropic,
homogeneous situation, with periodic boundary conditions. In this
case where the backscatter is secondary, we have shown that our model
still perform slightly better than traditional LES approach used in
this situation, based on eddy-viscosity concept. It would now be
interesting to test our model in more realistic situations, with
boundaries, where the stochastic modeling will become an essential
ingredient. Given that our model reduces to the ADM model in a
well-defined limit, we are confident that it can perform as well as
this last model. The interesting question is whether our dynamical
procedure increases its performance, and overcome certain limitations
of the ADM model. This will be the subject of a future work.

\section*{Acknowledgments}
This work benefited from computer time allocated by IDRIS run by
French CNRS and by CRIHAN (Centre de Ressource en Informatique
de Haute Normandie, France) We thank the ``Programme National de Plan\'etologie'',
the GDR Turbulence and the GDR Dynamo for support. We have benefited from numerous
discussion with our colleagues from GIT and from Lille University.



\end{document}